%% file: p206.tex
\documentclass[11pt,english,epsf]{article}%{scrartcl}
\usepackage{lmodern}
\usepackage[T1]{fontenc}
\usepackage[latin9]{inputenc}
\usepackage{geometry}
\usepackage{amsmath}
\usepackage{amssymb}
\usepackage{graphicx}
\newtheorem{theorem}{Theorem}

\newcommand {\dfn} {\stackrel{\Delta} {=}}
\newcommand {\exe} {\stackrel{\cdot} {=}}
\newcommand {\lexe} {\stackrel{\cdot} {\le}}

\newcommand {\hP} {\hat{P}}
\newcommand {\hH} {\hat{H}}

\newcommand {\bu} {\mbox{\boldmath $u$}}
\newcommand {\bv} {\mbox{\boldmath $v$}}

\newcommand {\bx} {\mbox{\boldmath $x$}}
\newcommand {\by} {\mbox{\boldmath $y$}}

\newcommand {\bE} {\mbox{\boldmath $E$}}

\newcommand {\bU} {\mbox{\boldmath $U$}}

\newcommand {\tW} {\tilde{W}}
\newcommand {\bX} {\mbox{\boldmath $X$}}

\newcommand{\calC}{{\cal C}}

\newcommand{\calE}{{\cal E}}

\newcommand{\calI}{{\cal I}}

\newcommand{\calP}{{\cal P}}

\newcommand{\calS}{{\cal S}}
\newcommand{\calT}{{\cal T}}
\newcommand{\calU}{{\cal U}}

\newcommand{\calX}{{\cal X}}
\newcommand{\calY}{{\cal Y}}

\allowdisplaybreaks

\begin{document}
\thispagestyle{empty}
\title{Error Exponents of Typical Random Trellis Codes
\thanks{This research was supported by the Israel Science Foundation (ISF),
grant no.\ 137/18.}
}
\author{Neri Merhav}
\date{}
\maketitle

\begin{center}
The Andrew \& Erna Viterbi Faculty of Electrical Engineering\\
Technion - Israel Institute of Technology \\
Technion City, Haifa 32000, ISRAEL \\
E--mail: {\tt merhav@ee.technion.ac.il}\\
\end{center}
\vspace{1.5\baselineskip}
\setlength{\baselineskip}{1.5\baselineskip}

\begin{abstract}
In continuation to an earlier work, where error exponents of typical random codes
were studied in the context of general block coding, with no underlying
structure, here we carry out a parallel study on typical random,
time--varying trellis codes for general discrete memoryless channels, 
focusing on a certain range of low rates.
By analyzing an upper bound to the error probability of the typical
random trellis code, using the method of types,
we first derive a Csisz\'ar--style error exponent formula (with respect to
the constraint length), which allows to easily
identify and characterize properties of good codes and dominant error events.
We also derive a Gallager--style form of this error exponent, which turns out
to be related to the expurgated error exponent.
The main result is further extended to
channels with memory and mismatch.\\

\end{abstract}

\newpage
\section{Introduction}

Following the work of Barg and Forney \cite{BF02}, Nazari
\cite{Nazari11} and Nazari {\it et al.} \cite{NAP14}, in a recent work
\cite{trc}, the error exponent of the typical random block code
for a general discrete memoryless channel (DMC) was studied. The error
exponent of the typical random code (TRC) was defined as the long--block limit of the
negative normalized {\it expectation of the logarithm} of the error probability, as
opposed to the classical random coding exponent, defined as the
negative normalized {\it logarithm of the expectation} of the error probability.
The investigation of error exponents for TRCs
was motivated in \cite[Introduction]{trc} by a few points: (i) Owing to Jensen's
inequality, it cannot be smaller than the random coding error exponent, and so,
it is a more optimistic
performance measure than the ordinary random coding exponent, especially
at low rates. (ii) Given that a certain measure concentration
property holds, it is more relevant as a performance
metric, since the code is normally assumed to be randomly selected just once, and
then used repeatedly. (iii) It captures correctly the behavior of
random--like codes \cite{Battail95},
which are well known to be very good codes. 

In \cite{trc},
an exact single--letter expression was derived for the error exponent function of the
TRC assuming a general discrete memoryless channel (DMC) and an
ensemble of fixed composition codes. Among other things, it was shown
in \cite{trc} (similarly as in \cite{BF02} and \cite{Nazari11}),
that the TRC error exponent is: (i) the same as the expurgated exponent at zero
rate, (ii) below the expurgated exponent, but
above the random coding exponent for low positive rates, and (iii) the same as 
the random coding exponent beyond a certain rate.

In view of the practical importance and the rich literature 
on trellis codes, and convolutional codes in 
particular (see, e.g., \cite{Costello74}, \cite{Forney74}, \cite{Johannesson77},
\cite{JR99}, \cite{Massey76}, \cite{SF00}, \cite{VO69}, \cite{VO79},
\cite{ZSSHJ99} just to name a few, as well as
and many references therein),
the purpose of this paper is to study the behavior and the performance of 
typical random trellis codes.
More specifically, our aim is at an investigation 
parallel to that of \cite{trc}, in the realm of 
ensembles of time--varying trellis codes. 
The main motivation is to compare the error 
exponent of the typical random trellis code
to that of the typical block code on the basis of similar 
decoding complexity, in the spirit 
of the similar comparison in \cite[Chap.\ 5]{VO79}, which was
carried out for the ordinary random coding exponents of the two classes of codes.
Technically speaking, our main result is that the error exponent of the 
typical random, time--varying trellis code is lower bounded by 
a certain expression that is related to the 
expurgated exponent, and its value lies between
those of the convolutional random coding error exponent and the
convolutional--coding 
expurgated exponent functions \cite{VO69}, \cite[Sect.\ 5]{VO79}. 
For the subclass of linear trellis codes, namely, time--varying convolutional
codes, the result is improved: the typical 
time--varying convolutional code achieves the convolutional--coding expurgated
exponent, provided that the channel is binary--input, output--symmetric (see
also \cite{VO69}).
In other words, in the 
limit of large constraint length, 
a randomly selected time--varying convolutional code achieves the
convolutional expurgated exponent with an overwhelmingly high probability.
This is parallel to a similar behavior in the context of 
ordinary random block codes (without structure), where the error exponent of
the typical random code is inferior to the corresponding expurgated exponent,
and superior to the random coding error exponent (at low rates), but when it
comes to linear random codes, the typical--code error exponent coincides with 
the expurgated exponent.

These results both sharpen and generalize some
earlier statements on the fraction of time--varying (or periodically time--varying) 
convolutional codes with certain 
properties (see, for example, 
\cite[Lemma 3.33, Lemma 4.15]{JR99}), and in particular, the fact 
that (at least) half of the convolutional codes achieve 
the convolutional coding exponent \cite[Theorem]{VO69}. Beyond this, 
our contributions are in several aspects.
\begin{enumerate}
\item Our analysis provides a fairly clear insight on the behavior
of the typical codes, i.e., their free distances and their distance enumerators.
\item Thanks to the use the method of types, we are able to characterize
the dominant error events, that is, typical lengths of error bursts and
joint types of incorrect trellis paths together with the correct path, which
are even more informative than distances.
\item Our analysis is considerably general: we address
general trellis codes (not merely convolutional codes) with a general random coding
distribution (not necessarily the uniform distribution) 
and a general discrete memoryless channel (DMC), not merely
binary--input, output--symmetric channels.
\item We further extend the results in two directions simultaneously, allowing
both channels with input memory and mismatch.
\end{enumerate}

The outline of the remaining part of this paper is the following.
In Section 2, we establish notation conventions, define the problem
setting, provide some background, and spell out the objectives
of the paper more formally. In Section 3, we state the main result, and in
Section 4 we prove it. Section 5 is devoted to some discussion, and finally,
in Section 6, we extend the main result to channels with memory and mismatch.

\section{Notation, Problem Setting, Background and Objectives}
\label{npbo}

\subsection{Notation}
\label{not}

Throughout the paper, random variables will be denoted by capital
letters, specific values they may take will be denoted by the
corresponding lower case letters, and their alphabets
will be denoted by calligraphic letters. Random
vectors and their realizations will be denoted,
respectively, by capital letters and the corresponding lower case letters,
both in the bold face font. Their alphabets will be superscripted by their
dimensions. For example, the random vector $\bX=(X_1,\ldots,X_r)$, ($r$ --
positive integer) may take a specific vector value $\bx=(x_1,\ldots,x_r)$
in $\calX^r$, the $r$--th order Cartesian power of $\calX$, which is
the alphabet of each component of this vector.
The probability of an event $\calE$ will be denoted by $\mbox{Pr}\{\calE\}$,
and the expectation
operator will be
denoted by
$\bE\{\cdot\}$. For two
positive sequences $\{a_k\}$ and $\{b_k\}$, the notation $a_k\exe b_k$ will
stand for equality in the exponential scale, that is,
$\lim_{k\to\infty}\frac{1}{k}\log \frac{a_k}{b_k}=0$. Similarly,
$a_k\lexe b_k$ means that
$\limsup_{k\to\infty}\frac{1}{k}\log \frac{a_k}{b_k}\le 0$, and so on.
The indicator function
of an event $\calE$ will be denoted by $\calI\{E\}$. 

The empirical distribution of a string of symbols in a finite alphabet $\calX$,
denoted by $\hat{P}_X$, is the vector of relative frequencies
$\hat{P}_X(x)$
of each symbol $x\in\calX$ along the string. Here $X$ denotes an auxiliary random
variable (RV) distributed according to this distribution. 
Information measures associated with empirical distributions
will be denoted with `hats'.
For example, the entropy associated with
the empirical distribution $\hat{P}_X$, namely, the empirical entropy, will be denoted by
$\hat{H}(X)$. 
Similar conventions will apply to the joint empirical
distribution, the joint type class, the conditional empirical distributions
and the conditional type classes associated with pairs (and multiples) of
sequences of length $r$.
Accordingly, $\hP_{XX^\prime}$ will be the joint empirical
distribution associated with a pair of strings of the same length,
$\hH(X,X^\prime)$ will designate the empirical joint entropy,
and $\hH(X|X^\prime)$ will be the empirical conditional entropy.

\subsection{Problem Setting}
\label{ps}

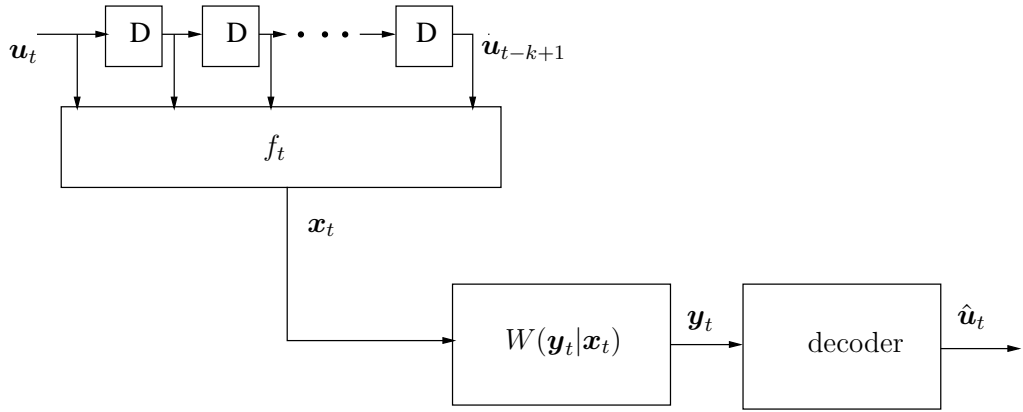
\begin{figure}[ht]
\vspace*{1cm}
\hspace*{2cm}\input{trelliscode.pstex_t}
\caption{\small Block diagram of a communication system based on a 
time--varying trellis code.}
\label{blockdiagram}
\end{figure}

Consider the system configuration depicted in Fig.\ \ref{blockdiagram}.
Let the information source, $U_1, U_2,\ldots$, be the binary symmetric source
(BSS), i.e., an infinite sequence of binary random variables taking on values in
$\calU=\{0,1\}$, independently of each other, and with equal probabilities for
`0' and `1'. We shall group the bits of this information source in blocks of
length $m$, and denote $\bU_t=(U_{m(t-1)+1},U_{m(t-1)+2},\ldots,U_{mt})$,
$\bU_t\in\calU^m$, $t=1,2,\ldots$.

A {\it time--varying trellis code} of rate $R=m/n$ and with memory size $k$, 
is a sequence of functions $f_1,
f_2,\ldots$, $f_t:\calU^{mk}\to\calX^n$, 
$t=1,2,\ldots$, where $\calX$ is the finite channel input
alphabet of size $J$. 
When fed with an input information sequence,
$\bu_1,\bu_2,\ldots$, which is a realization of $\bU_1,\bU_2,\ldots$, the
time--varying trellis codes outputs a code sequence, $\bx_1,\bx_2,\ldots$, according to
\begin{equation}
\bx_t=f_t(\bu_t,\bu_{t-1},\ldots,\bu_{t-k+1}),~~~~~~~t=1,2,\ldots
\end{equation}
The product $mk$ designates the {\it constraint length} of the trellis code, and it
will henceforth be denoted 
by $K$. As is well known, a trellis code is a special case of
a finite--state encoder whose total number of states is $2^K$. 
On the other hand,
a convolutional code is a special case of a trellis code where $\{f_t\}$ are
linear functions over the relevant field.

A discrete memoryless channel (DMC) $W$ is defined by a set of single--letter conditional
probabilities (or probability density functions),
$\{W(y|x),~x\in\calX,~y\in\calY\}$, where $\calX$ is as before and $\calY$ is
the channel output alphabet, which may be discrete or
continuous.\footnote{Throughout the sequel, we will treat $\calY$ as a
discrete alphabet, with the understanding that in the continuous case, all
summations over $\calY$ should be replaced by integrals.}
When the
channel is fed by a sequence, $x_1,x_2,\ldots$, $x_t\in\calX$, $t=1,2,\ldots$
(a realization of a random process, $X_1,X_2,\ldots$),
it responds by generating a
corresponding output sequence, $y_1,y_2,\ldots$, $y_t\in\calY$,
$t=1,2,\ldots$ (a realization of a random process, $Y_1,Y_2,\ldots$), according to
\begin{equation}
\label{channel}
\mbox{Pr}\{Y_1=y_1,Y_2=y_2,\ldots,Y_r=y_r|X_1=x_1,X_2=x_2,\ldots,X_r=x_r\}=
\prod_{t=1}^rW(y_t|x_t).
\end{equation}

As customary, we assume that the trellis code is decoded in long blocks using the
maximum--likelihood (ML) decoder, which is implementable by the Viterbi algorithm, and by
terminating each block with $m(k-1)$ zero input bits in order to reset the state of
the encoder. As mentioned earlier, we also extend the results to
channels with input memory
(inter-symbol interference) along with
mismatched decoding metrics, which are still implementable by the
Viterbi Algorithm.

We consider the ensemble of time--varying trellis codes where for every
$t=1,2,\ldots$ and every possible value of 
$(\bu_t,\bu_{t-1},\ldots,\bu_{t-k+1})\in\calU^K)$,
the value of $f_t(\bu_t,\bu_{t-1},\ldots,\bu_{t-k+1})\in\calX^n$ is
selected independently at random under the i.i.d.\ distribution $Q^n$, namely,
each one of the $n$ components of
$f_t(\bu_t,\bu_{t-1},\ldots,u_{t-k+1})\in\calX^n$ is randomly drawn independently
under a fixed distribution $Q$ over $\calX$. For the case of time--varying
convolutional codes, the symbols $\{x_t\}$ are assumed binary ($J=2$), and
$\{f_t\}$ are assumed linear functions over $\mbox{GF}(2)$,
namely,
\begin{equation}
f_t(\bu_t,\ldots,\bu_{t-k+1})=\bx_{0,t}\oplus\sum_{j=0}^{k-1}\bu_{t-j}G_j(t),
\end{equation}
where $\{\bu_{t-j}\}$ are considered row--vectors of dimension $m$,
$\{\bx_{0,t}\}$ are binary vectors of dimension $n$,
$\{G_j(t)\}$ are binary $m\times n$ matrices, the operations $\oplus$ and $\sum$
both designate summations modulo 2,
and the channel is assumed binary--input, output--symmetric.
The entries of $\{\bx_{0,t}\}$ and $\{G_j(t)\}$ are randomly
and independently selected with equal probabilities of $0$ and $1$.

\subsection{Background}
\label{bg}

The traditional ensemble performance metric is the exponential decay rate
(as a function of $K$) of the expectation of the first--error event 
probability, or the per--node error probability \cite[p.\ 243]{VO79}, as well
as the
related bit error probability, 
\begin{equation}
\calE_{\mbox{\tiny rtc}}(R,Q)=\liminf_{K\to\infty}
\left\{-\frac{\log \bE P_{\mbox{\tiny e}}}{K}\right\},
\end{equation}
where the subscript ``rtc'' stands for ``random trellis code''
and accordingly, the expectation is w.r.t.\ the
randomness of the time--varying trellis code, see, e.g., \cite[Chap.\ 5]{VO79}. 
As shown in \cite[Sect.\ 5.1]{VO79}, the result for random time--varying convolutional
codes, which easily extends
to random time---varying trellis codes,
is that this error exponent is essentially\footnote{The actual exponent is
slightly smaller than that, but
by an amount $\epsilon$ that can be made arbitrarily small. Here and in the
sequel, we
will ignore this very small loss.}
given by 
\begin{equation}
\calE_{\mbox{\tiny rtc}}(R,Q)\ge E_{\mbox{\tiny rtc}}(R,Q)\dfn 
\left\{\begin{array}{ll}
R_0(Q)/R & R < R_0(Q)\\
E_0(\rho_{\mbox{\tiny rtc}}(R),Q)/R & R > R_0(Q)\end{array}\right.
\end{equation}
where $\rho_{\mbox{\tiny rtc}}(R)$ is the solution $\rho$
the equation $R = E_0(\rho,Q)/\rho$, $E_0(\rho,Q)$ being the Gallager
function,
\begin{equation}
E_0(\rho,Q)=-\log\left(\sum_y\left[\sum_x
Q(x)W(y|x)^{1/(1+\rho)}\right]^{1+\rho}\right),
\end{equation}
and $R_0(Q)=E_0(1,Q)$. The best result is obtained, of course, upon maximizing
over $Q$, in which case, for $R > R_0=\max_Q R_0(Q)$, the resulting 
error exponent is the best achievable error exponent,
as it meets the converse bound of \cite[Theorem 5.4.1]{VO79}.\footnote{Although
the converse bound in \cite[Sect.\ 5.4]{VO79} is proved with convolutional
codes in mind, the
linearity of convolutional codes is not really used there, and so the very same
proof applies also to non--linear trellis codes.} It follows then that there
is room for improvement only for rates below $R_0$. Indeed, an improvement in
this range is
accomplished, for binary--input, output symmetric channels \cite[p.\ 86]{VO79}, 
by an expurgated
bound, derived in \cite{VO69}, \cite[Sect.\ 5.3]{VO79}, and given by
\begin{equation}
\label{cex}
E_{\mbox{\tiny cex}}(R,Q)=\frac{E_{\mbox{\tiny x}}(\rho_{\mbox{\tiny cex}}(R),Q)}{R}, 
\end{equation}
where $\rho_{\mbox{\tiny cex}}(R)$ is the solution $\rho\ge
1$ to the
equation $R = E_{\mbox{\tiny x}}(\rho,Q)/\rho$, with $E_{\mbox{\tiny
x}}(\rho,Q)$ being defined as
\begin{equation}
E_{\mbox{\tiny x}}(\rho,Q)=-\rho\log\left[\sum_{x,x^\prime}Q(x)Q(x^\prime)
\left(\sum_y\sqrt{W(y|x)W(y|x^\prime)}\right)^{1/\rho}\right].
\end{equation}
More precisely, in \cite{VO69} the main theorem asserts that for {\it at least
half} of
the rate--$1/n$ time--varying convolutional codes, the probability of error
does not exceed 
\begin{equation}
\label{vo69}
\left(\frac{2L}{1-2^{-\epsilon/\rho
R}}\right)^\rho\cdot\exp\{-KE_{\mbox{\tiny cex}}(\rho,Q)\},
\end{equation}
where $Q$ is the binary symmetric source (which in our notation, means
the uniform distribution over
the binary alphabet $\calX$), 
$L$ is the block length, $\epsilon = E_{\mbox{\tiny x}}(\rho,Q)-\rho R > 0$ 
is an arbitrarily small positive
real, and $\rho\ge 1$ is any number that satisfies $R= [E_{\mbox{\tiny
x}}(\rho,Q)-\epsilon]/\rho < R_0$. It is clear from the proof of this theorem
that choosing to refer to exactly half of the codes is quite arbitrary, and a similar
bound, with the same exponential rate (assuming that $L$ is sub--exponential
in $K$), would apply to any, arbitrarily large,
fraction of the codes, at the expense of increasing the pre--exponential
factor of (\ref{vo69}) accordingly. For example, if the factor $2L$ at the numerator of the
pre--exponent of (\ref{vo69}) is replaced by $100L$, then the bound would apply to at
least 99\% of the
time--varying convolutional codes with block length $L$, and so on.
This indicates that the ensemble of convolutional 
codes obeys a {\it measure concentration
property} concerning their error exponent.\footnote{As mentioned in the Introduction,
several assertions in the same spirit can be found also in \cite{JR99}, see
for example,
Lemmas 3.33 and 4.15 therein.}

\subsection{Objectives}
\label{obj}

The purpose of this work is to study the above 
mentioned measure concentration property in
a systematic manner and to broaden the scope in several directions at
the same time, as will be specified shortly. 
In this context, similarly as in \cite{trc}, we refer to the error
exponent of typical random trellis code, and as discussed in \cite[Introduction]{trc}, 
if the ensemble of codes possesses the relevant measure concentration
property associated with exponential error bounds, then the error exponent of
the the {\it typical random trellis code}, 
is captured by the quantity
\begin{equation}
\calE_{\mbox{\tiny trtc}}(R,Q)\dfn\liminf_{K\to\infty}
\left\{-\frac{\bE\log P_{\mbox{\tiny e}}}{K}\right\},
\end{equation}
which is similar to the above definition of $\calE_{\mbox{\tiny rtc}}(R,Q)$, except that
the expectation operator and the logarithmic function are commuted. It will be
understood that the limit of $K=mk\to\infty$ will be taken under the regime
where $m$ and $n$ (and hence also $R=m/n$) are held fixed whereas $k\to\infty$.
A similar definition will apply to the smaller ensemble to time--varying
convolutional codes and it will be denoted by $\calE_{\mbox{\tiny
trcc}}(R,Q)$, where the subscript stands for {\it typical random convolutional
code}.

\section{Main Result}

Our main theorem has two parts, where the second part
actually follows directly from \cite{VO69} (as discussed in Subsection
\ref{bg}) and is included here for completeness.
\begin{theorem}
\label{thm}
Consider the problem setting defined in Subsection \ref{ps}. Then,
for $R < R_0(Q)$,
\begin{enumerate}
\item[(a)]
\begin{equation}
\calE_{\mbox{\tiny trtc}}(R,Q)\ge E_{\mbox{\tiny trtc}}(R,Q)\dfn
\frac{E_{\mbox{\tiny x}}(\rho_{\mbox{\tiny trtc}}(R),Q)}{R},
\end{equation}
where $\rho_{\mbox{\tiny trtc}}(R)$ is the solution, $\rho\ge 1$, to the equation
\begin{equation}
R=\frac{E_{\mbox{\tiny x}}(\rho,Q)}{2\rho-1}.
\end{equation}
\item[(b)]
For the ensemble of time--varying convolutional 
codes and the binary--input output symmetric
channel (with $Q(0)=Q(1)=\frac{1}{2}$),
\begin{equation}
\calE_{\mbox{\tiny trcc}}(R,Q)\ge E_{\mbox{\tiny cex}}(R,Q).
\end{equation}
\end{enumerate}
\end{theorem}

We emphasize that here the setup is considerably extended relative to that of
\cite{VO69}, especially in part (a). This extension takes place in
several dimensions at the same time:
\begin{enumerate}
\item Allowing general rational coding rates, $R=m/n$, rather than 
$R=1/n$.
\item Using ensembles with a general random coding distribution $Q$,
instead of just the 
uniform distribution. In this case, assertions about fractions of codes with
certain properties are replaced by parallel assertions concerning 
(high) probabilities of possessing
these properties.
\item Assuming a general DMC, 
not necessarily a binary--input, output symmetric
channel.
\item As was mentioned already, we are referring to general trellis codes, as an
extension to convolutional codes, which are linear.
\item A further extension is for mismatched decoding for a channel with input memory.
\end{enumerate}
Furthermore, our analysis, which is strongly based
on the method of types, will provide some insights on the character of 
two ingredients of interest:
\begin{enumerate}
\item Structure and distance enumeration (or more generally, type class
enumeration) of the typical random trellis code, that achieves the
convolutional coding expurgated exponent.
\item Error events that dominate the error probability: joint types of
decoded trellis paths and the correct paths, along with the lengths of the
typical error bursts.
\end{enumerate}
These points, among others, will be discussed in mode detail in
Section \ref{dis}.

\section{Proof of Theorem \ref{thm}}
\label{proof}

Here we prove part (a) only, because part (b) can be obtained in a very similar manner
by a small modification in a few places. Also, as discussed in Subsection
\ref{bg}, part
(b) was actually proved already in \cite{VO69} (at
least for rate--$1/n$ codes, but the extension to $m/n$--codes is not
difficult).

Clearly, in order to derive a bound on $\calE_{\mbox{\tiny trtc}}(R,Q)$, 
we have to assess $\bE\log P_{\mbox{\tiny e}}(\calC_k)$,
where $\calC_k$ designates a randomly selected trellis code with memory $k$
(and constraint length $K=mk$) in the ensemble described in Subsection
\ref{ps}. Our first observation is the following: 
suppose we can define, for every $k\ge 1$, a
subset $\calT_k$ of codes $\{\calC_k\}$ whose probability, $1-\epsilon_k\dfn
\mbox{Pr}\{\calT_k\}$, tends to unity
as $k\to\infty$. Then,
\begin{eqnarray}
\label{condexp}
\bE\log P_{\mbox{\tiny e}}(\calC_k)&=&\mbox{Pr}\{\calT_k
\}\cdot\bE\{\log P_{\mbox{\tiny e}}(\calC_k)|\calC_k\in
\calT_k\}+\mbox{Pr}\{\calT_k^{\mbox{\tiny c}}\}\cdot\bE\{\log P_{\mbox{\tiny
e}}(\calC_k)|\calC_k\in \calT_k^{\mbox{\tiny c}}\}\nonumber\\
&\le&(1-\epsilon_k)\cdot\bE\{\log P_{\mbox{\tiny e}}(\calC_k)|\calC_k\in\calT_k
\}+\epsilon_k\cdot\log 1\nonumber\\
&=&(1-\epsilon_k)\cdot\bE\{\log P_{\mbox{\tiny
e}}(\calC_k)|\calC_k\in\calT_k\}\nonumber\\
&\le&(1-\epsilon_k)\cdot\log\left[\max_{\calC_k\in\calT_k} P_{\mbox{\tiny
e}}(\calC_k)\right].
\end{eqnarray}
Thus, if we can define a subset of codes $\calT_k$, which on the one hand, has very
high probability, and on the other hand, there is a uniform upper bound on
$P_{\mbox{\tiny e}}(\calC_k)$ for every $\calC_k\in\calT_k$, this would yield
a lower bound on the error exponent of the typical random trellis code.
We will use this simple observation shortly after we define the subset
$\calT_k$.

As mentioned earlier,
we are assuming that each transmitted block is terminated by $k-1$
all--zero input vectors (each of dimension $m$) in order to reset the state of
the shift
register of the trellis encoder. Similarly as in linear
convolutional codes, here too, every incorrect path $\{\bv_t\}$, diverging from the
correct path, $\{\bu_t\}$,
at a given node $j$ and re-merging with the correct path exactly after $k+\ell$
branches, must have the form
$$\bv_j,\bv_{j+1},\ldots,\bv_{j+\ell},\bu_{j+\ell+1},
\bu_{j+\ell+2},\ldots,\bu_{j+\ell+k-1},$$
where $\bv_j$ and $\bv_{j+\ell}$ can be any one of the $2^m-1$ incorrect input
$m$--vectors at nodes $j$ and $j+\ell$, respectively. Between $j$ and
$j+\ell$ there should be no sub-strings of $k-1$ consecutive correct inputs.
Thus, overall there are
no more than $(2^m-1)2^{m\ell}$ such incorrect paths \cite[p.\ 311]{VO79}.
Following a similar\footnote{Note that here, unlike in \cite{VO79},
in part (a) of Theorem \ref{thm},
we are considering general trellis codes, not convolutional codes, which are
linear. Therefore, we cannot assume, without loss of generality, that
the all--zero message was sent, but rather average over all input messages.
In part (b), on the other hand, this averaging is not needed. This difference 
causes certain modifications in the analysis, which
yield eventually $E_{\mbox{\tiny
cex}}(R,Q)$.} line of thought as in the
derivations of \cite{VO79}, for a given trellis code $\calC_k$, 
the probability of an error event beginning at any
given node is upper bounded by
\begin{equation}
P_{\mbox{\tiny e}}(\calC_k)\le \sum_{\ell\ge
1}\frac{1}{2^{m\ell}}\sum_{\bx\in\calX^{k+\ell}}\sum_{\bx^\prime\in\calX^{k+\ell}}
\mbox{Pr}\left\{W(\by|\bx^\prime)\ge W(\by|\bx)\right\},
\end{equation}
where $\bx$ designates the codeword associated with the correct path and
$\bx^\prime$ stands for any incorrect path diverging from the correct path at
node $j$ and re-merging at $j+k+\ell$. Since $\bx$ and $\bx^\prime$ may
disagree at no more than $n(k+\ell)$ channel uses, the summand is actually the pairwise
error probability associated with two vectors of length $n(k+\ell)$, and it
depends only on the joint empirical distribution of these two $n(k+\ell)$--vectors, which we
denote by $\hP_{XX^\prime}$. In particular, by the Chernoff bound, it is
readily seen that for a given pair $(\bx,\bx^\prime)$,
\begin{eqnarray}
\mbox{Pr}\left\{W(\by|\bx^\prime)\ge
W(\by|\bx)\right\}&\le&\exp\left\{-n(k+\ell)\max_{0\le s\le 1}
\sum_{x,x^\prime}P_{XX^\prime}(x,x^\prime)d_s(x,x^\prime)\right\}\nonumber\\
&\dfn&\exp_2\left\{-n(k+\ell)\max_{0\le s\le 1}\Delta_s(\hP_{XX^\prime})\right\}\nonumber\\
&\dfn&\exp_2\left\{-n(k+\ell)\Delta(\hP_{XX^\prime})\right\},
\end{eqnarray}
where
\begin{equation}
d_s(x,x^\prime)=-\log_2\left[\sum_y W^{1-s}(y|x)W^s(y|x^\prime)\right],
\end{equation}
is the Chernoff distance between $x$ and $x^\prime$.
It follows then that
\begin{equation}
\label{first}
P_{\mbox{\tiny e}}(\calC_k)\le \sum_{\ell\ge
1}2^{-m\ell}\sum_{\{\hP_{XX^\prime}\}}N_\ell(\hP_{XX^\prime})\cdot
\exp\left\{-n(k+\ell)\Delta(\hP_{XX^\prime})\right\},
\end{equation}
where $N_\ell(\hP_{XX^\prime})$ is the number of pairs
$\{(\bx,\bx^\prime)\}\in\calX^{2n(k+\ell)}$
having joint empirical distribution 
that is given by $\hP_{XX^\prime}$.
Here, the inner summation over $\{\hP_{XX^\prime}\}$ is defined over 
the set $\calP^{n(k+\ell)}$ of all possible
empirical distributions of pairs of vectors in $\calX^{n(k+\ell)}$.
For a given joint empirical distribution $\hP_{XX^\prime}$, we denote
\begin{equation}
D(\hP_{XX^\prime}\|Q\times
Q)=\sum_{x,x^\prime\in\calX}\hP_{XX^\prime}(x,x^\prime)\log_2\frac{\hP_{XX^\prime}(x,x^\prime)}
{Q(x)Q(x^\prime)}.
\end{equation}
We note that
\begin{eqnarray}
\bE\{N_\ell(\hP_{XX^\prime})\}&\le&
(2^m-1)2^{2m\ell}\cdot\mbox{Pr}\{(\bx,\bx^\prime)~\mbox{have joint
type}~\hP_{XX^\prime}\}\nonumber\\
&\le&(2^m-1)2^{2m\ell}\cdot\exp_2\{-n(k+\ell)D(\hP_{XX^\prime}\|Q\times
Q)\}\nonumber\\
&=&(2^m-1)\cdot\exp_2\left\{m[2\ell-(k+\ell)D(\hP_{XX^\prime}\|Q\times
Q)/R]\right\}.
\end{eqnarray}
We now define
$\calT_k$ as the subset of codes, henceforth referred to as the {\it typical
trellis codes}, with the following property for a given
arbitrarily small $\epsilon > 0$:
for every $\ell\ge 1$ and every empirical joint distribution $\hP_{XX^\prime}$
derived from $n(k+\ell)$--vectors: 
\begin{itemize}
\item $N_\ell(\hP_{XX^\prime})=0$ whenever
$\bE\{N_\ell(\hP_{XX^\prime})\} < (2^m-1)\cdot 2^{-n(k+\ell)\epsilon}$, and 
\item $N_\ell(\hP_{XX^\prime})\le
2^{n(k+\ell)\epsilon}\cdot \bE\{N_\ell(\hP_{XX^\prime})\}$ whenever
$\bE\{N_\ell(\hP_{XX^\prime})\} \ge (2^m-1)\cdot 2^{-n(k+\ell)\epsilon}$. 
\end{itemize}
Obviously, by the Markov inequality, for every $\ell$ and $\hP_{XX^\prime}$ in
the first category, we have
\begin{equation}
\mbox{Pr}\{N_\ell(\hP_{XX^\prime})\ge 1\}\le \bE\{N_\ell(\hP_{XX^\prime})\} <
(2^m-1)\cdot 2^{-n(k+\ell)\epsilon},
\end{equation}
and similarly, for $\ell$ and $\hP_{XX^\prime}$ in the second category,
we have
\begin{equation}
\mbox{Pr}\{N_\ell(\hP_{XX^\prime})> 2^{n(k+\ell)\epsilon}\cdot
\bE\{N_\ell(\hP_{XX^\prime})\}\le
2^{-n(k+\ell)\epsilon} < (2^m-1)\cdot 2^{-n(k+\ell)\epsilon}.
\end{equation}
It follows by the union bound that
\begin{eqnarray}
\label{PGkc}
\mbox{Pr}\{\calT_k^{\mbox{\tiny c}}\}&\le&(2^m-1)\sum_{\ell\ge
1}\sum_{\{\hP_{XX^\prime}\}} 2^{-n(k+\ell)\epsilon}\nonumber\\
&\le&(2^m-1)\sum_{\ell\ge 1}[n(k+\ell)+1)^{J^2}\cdot 2^{-n(k+\ell)\epsilon}\nonumber\\
&=&(2^m-1)\cdot \sum_{\ell\ge k+1}(n\ell+1)^{J^2}\cdot 2^{-n\ell\epsilon}\nonumber\\
&=&(2^m-1)\cdot \sum_{\ell\ge k+1}
\exp_2\left\{-n\ell\left[\epsilon-\frac{J^2\log(n\ell+1)}{n\ell}\right]\right\}.
\end{eqnarray}
The sequence $\{\frac{\log(n\ell+1)}{n\ell}\}$ is monotonically decreasing 
and so, since $\ell\ge k+1$, we have, for large enough $k$,
$$\frac{J^2\log(n\ell+1)}{n\ell}\le
\frac{J^2\log[n(k+1)+1]}{n(k+1)}\le \frac{\epsilon}{2},$$
and then the last line of (\ref{PGkc})
cannot exceed the sum of the geometric series,
$(2^m-1)\cdot 2^{-n(k+1)\epsilon/2}/(1-2^{-n\epsilon/2})$,
which tends to zero as $k\to\infty$. Thus, $\mbox{Pr}\{\calT_k\}$ tends to
unity as $k\to\infty$. 
Denoting
\begin{eqnarray}
\calS_\ell^\prime&=&\{\hP_{XX^\prime}\in\calP^{n(k+\ell)}:~\bE\{N_\ell(\hP_{XX^\prime})\}
\ge (2^m-1)\cdot 2^{-n(k+\ell)\epsilon}\}\nonumber\\
&\subseteq&\left\{\hP_{XX^\prime})\in\calP^{n(k+\ell)}:~2\ell \ge
\frac{k+\ell}{R}[D(\hP_{XX^\prime}\|Q\times
Q)-\epsilon]\right\}\nonumber\\
&\dfn&\calS_\ell,
\end{eqnarray}
it now follows that for every typical trellis code, $\calC_k\in\calT_k$,
\begin{eqnarray}
P_{\mbox{\tiny e}}(\calC_k)&\le&\sum_{\ell\ge
1}2^{-m\ell}\sum_{\{\hP_{XX^\prime})\in\calS_\ell^\prime\}}
N_{\ell}(\hP_{XX^\prime})\cdot\exp_2\{-n(k+\ell)\Delta(\hP_{XX^\prime})\}\nonumber\\
&\le&(2^m-1)\sum_{\ell\ge 1}\sum_{\{\hP_{XX^\prime})\in\calS_\ell\}}
\exp_2\{m(\ell-(k+\ell)[D(\hP_{XX^\prime}\|Q\times
Q)+\nonumber\\
& &\Delta(\hP_{XX^\prime})-\epsilon]/R)\}.
\end{eqnarray}
In order to address this summation over $\calS_\ell$, let us partition it
as the disjoint union of the subsets
\begin{equation}
\calS_{\ell,i}=\calS_\ell\cap\{\hP_{XX^\prime}\in\calP^{n(k+\ell)}:~R_{i-1}\le
D(\hP_{XX^\prime}\|Q\times Q)< R_i\},~~~~R_i=i\epsilon,~~i=1,2,\ldots,\lceil
2R/\epsilon\rceil
\end{equation}
and observe that for a given $i$, $\calS_{\ell,i}$ is non--empty only when
$2\ell \ge (k+\ell)(R_{i-1}-\epsilon)/R$, or equivalently,
$$\ell\ge \frac{k(R_{i-1}-\epsilon)}{2R-R_{i-1}+\epsilon}\dfn
k\theta(R_{i-1}).$$
Then,
\begin{eqnarray}
\label{bound}
P_{\mbox{\tiny e}}(\calC_k)&\le&\sum_{i=1}^{\lceil 2R/\epsilon\rceil}
\sum_{\ell\ge 1}\sum_{\{\hP_{XX^\prime}\in\calS_{\ell,i}\}}
\exp_2\{m[\ell-(k+\ell)[D(\hP_{XX^\prime}\|Q\times
Q)+\Delta(\hP_{XX^\prime})-\epsilon]/R\}\nonumber\\
&\le&\sum_{i=1}^{\lceil 2R/\epsilon\rceil}\sum_{\ell\ge k\theta(R_{i-1})}
[n(k+\ell)+1]^{J^2}\max_{\{\hP_{XX^\prime}:~D(\hP_{XX^\prime}\|Q\times Q)\le
R_i\}}\nonumber\\
& &\exp_2\{m[\ell-(k+\ell)[R_{i-1}+\Delta(\hP_{XX^\prime})-\epsilon]/R\}\nonumber\\
&=&\sum_{i=1}^{\lceil R/\epsilon\rceil}\sum_{\ell\ge k\theta(R_{i-1})}
[n(k+\ell)+1]^{J^2}
\exp_2\{m[\ell-(k+\ell)[R_{i-1}+Z(R_i)-\epsilon]/R\}\nonumber\\
&=&\sum_{i=1}^{\lceil
R/\epsilon\rceil}\exp_2\{-K[R_{i-1}+Z(R_i)-\epsilon]/R\}\times\nonumber\\
& &\sum_{\ell\ge k\theta(R_{i-1})}[n(k+\ell)+1]^{J^2}
\exp_2\{-m\ell[R_{i-1}+Z(R_i)-R-\epsilon]/R\},
\end{eqnarray}
where we have defined
\begin{equation}
Z(R_i)=\min\{\Delta(\hP_{XX^\prime}):~D(\hP_{XX^\prime}\|Q\times Q)\le R_i\}.
\end{equation}
Now observe that
\begin{eqnarray}
R_{i-1}+Z(R_i)&=&R_i+Z(R_i)-\epsilon\nonumber\\
&=&R_i+\min\{\Delta(\hP_{XX^\prime}):~D(\hP_{XX^\prime}\|Q\times Q)\le
R_i\}-\epsilon\nonumber\\
&\ge&\min_{\{\hP_{XX^\prime}:~D(\hP_{XX^\prime}\|Q\times Q)\le R_i\}}[
D(\hP_{XX^\prime}\|Q\times Q)+\Delta(\hP_{XX^\prime})]-\epsilon\nonumber\\
&\ge&\min_{\hP_{XX^\prime}}[
D(\hP_{XX^\prime}\|Q\times Q)+\Delta(\hP_{XX^\prime})]-\epsilon\nonumber\\
&=&\min_{\hat{P}_{XX^\prime}}\max_{0\le s\le
1}\left[D(\hat{P}_{XX^\prime}\|Q\times Q)+
\sum_{x,x^\prime}\hat{P}_{XX^\prime}(x,x^\prime)d_s(x,x^\prime)\right]-\epsilon\nonumber\\
&=&\max_{0\le s\le 1}\min_{\hat{P}_{XX^\prime}}
\left[D(\hat{P}_{XX^\prime}\|Q\times Q)+
\sum_{x,x^\prime}\hat{P}_{XX^\prime}(x,x^\prime)d_s(x,x^\prime)\right]-\epsilon\nonumber\\
&=&\max_{0\le s\le
1}\left\{-\log\left[\sum_{x,x'}Q(x)Q(x')2^{-d_s(x,x')}\right]\right\}-\epsilon\nonumber\\
&=&-\min_{0\le s\le
1}\log\left[\sum_{x,x',y}Q(x)Q(x')W^s(y|x)W^{1-s}(y|x')\right]-\epsilon\nonumber\\
&=&-\log\left[\sum_{x,x',y}Q(x)Q(x')\sqrt{W(y|x)W(y|x')}\right]-\epsilon\nonumber\\
&=&-\log\left(\sum_{y}\left[\sum_xQ(x)\sqrt{W(y|x)}\right]^2\right)-\epsilon\nonumber\\
&=&R_0(Q)-\epsilon,
\end{eqnarray}
where the commutation of the minimization and the maximization is allowed
by convexity--concavity of the objective, and the final minimization over $s$
is achieved by $s=1/2$ due to the convexity and the symmetry of the function
$\sum_{x,x',y}Q(x)Q(x')W^s(y|x)W^{1-s}(y|x')$ around $s=1/2$.
Thus, the series in the last line of (\ref{bound}) is convergent as long as $R <
R_0(Q)-2\epsilon$, and its exponential order as a function of $K$ (ignoring $\epsilon$--terms) 
is given by
\begin{eqnarray}
& &\frac{1}{R}\min_i\left\{R_i+Z(R_i)+\theta(R_i)[R_i+Z(R_i)-R]\right\}\nonumber\\
&=&\frac{1}{R}\min_i\left\{R_i+Z(R_i)+\frac{R_i}{2R-R_i}\cdot[R_i+Z(R_i)-R]\right\}\nonumber\\
&=&\min_i\frac{R_i+2Z(R_i)}{2R-R_i}\nonumber\\
&\ge&\inf_{\hat{R}< 2R}\frac{2Z(\hat{R})+\hat{R}}{2R-\hat{R}}\nonumber\\
&=&\inf_{\hat{R}/2 < R}\frac{Z(\hat{R})+\hat{R}/2}{R-\hat{R}/2}\nonumber\\
&=&\inf_{\hat{R} < R}\frac{Z(2\hat{R})+\hat{R}}{R-\hat{R}}\nonumber\\
&=&\inf_{\hat{R}< R}\inf_{\{P_{XX^\prime}:~
D(P_{XX^\prime}\|Q\times Q)\le 2\hat{R}\}}\max_{0\le s\le
1}\frac{\Delta_s(P_{XX^\prime})+\hat{R}}{R-\hat{R}}.
\end{eqnarray}
Thus, we have shown that the typical random trellis code error exponent
is lower bounded by
\begin{equation}
\label{csiszarstyle}
\calE_{\mbox{\tiny trtc}}(R,Q)\ge
\inf_{\hat{R}< R}\inf_{\{P_{XX^\prime}:~
D(P_{XX^\prime}\|Q\times Q)\le 2\hat{R}\}}\max_{0\le s\le
1}\frac{\Delta_s(P_{XX^\prime})+\hat{R}}{R-\hat{R}}.
\end{equation}
We next show that this expression is equivalent to the one asserted in part
(a) of Theorem
\ref{thm}. First, observe that since $\Delta_s(P_{XX^\prime})$ is a linear
functional of $P_{XX^\prime}$, then $\Delta(P_{XX^\prime})=
\max_{0\le s\le 1}\Delta(P_{XX^\prime})$
is convex in $P_{XX^\prime}$. We argue that the minimizer, $P_{XX^\prime}^*$, of
$\Delta(P_{XX^\prime})$ within the set $\{P_{XX^\prime}:~
D(P_{XX^\prime}\|Q\times Q)\le 2\hat{R}\}$ must be a symmetric distribution,
namely, $P_{XX^\prime}^*(x,x^\prime)=P_{XX^\prime}^*(x^\prime,x)$ for all
$x,x^\prime\in\calX$. To see why this is true, given any $P_{XX^\prime}$ that satisfies the
divergence constraint, define its transpose,
$\tilde{P}_{XX^\prime}$ by
$\tilde{P}_{XX^\prime}(x,x^\prime)=P_{XX^\prime}(x^\prime,x)$ for all
$x,x^\prime\in\calX$. Obviously,
$\Delta(\tilde{P}_{XX^\prime})=\Delta(P_{XX^\prime})$ because if $s^*$
achieves $\Delta(P_{XX^\prime})$, then $1-s^*$ achieves
$\Delta(\tilde{P}_{XX^\prime})$ and the value of the maximum is the same (just by swapping $x$
and $x^\prime$). Next, define
$\bar{P}_{XX^\prime}=\frac{1}{2}P_{XX^\prime}+\frac{1}{2}\tilde{P}_{XX^\prime}$. 
Then,
\begin{equation}
\Delta\left(\bar{P}_{XX^\prime}\right)=\Delta\left(\frac{1}{2}P_{XX^\prime}+
\frac{1}{2}\tilde{P}_{XX^\prime}\right)\le \frac{1}{2}\Delta(P_{XX^\prime})+
\frac{1}{2}\Delta(\tilde{P}_{XX^\prime})=\Delta(P_{XX^\prime}),
\end{equation}
and at the same time,
\begin{equation}
D(\bar{P}_{XX^\prime}\|Q\times Q)\le\frac{1}{2}D(P_{XX^\prime}\|Q\times Q)+
\frac{1}{2}D(\tilde{P}_{XX^\prime}\|Q\times Q)=D(P_{XX^\prime}\|Q\times
Q)\le 2\hat{R},
\end{equation}
so the divergence constraint is satisfied. It follows then that the symmetric
distribution $\bar{P}_{XX^\prime}$ is never worse than $P_{XX^\prime}$ in
terms of minimizing $\Delta(\cdot)$ under the divergence constraint. Thus, it
is sufficient to seek the minimizing $P_{XX^\prime}$ among the symmetric
distributions. However, given that $P_{XX^\prime}$ is symmetric, the maximizing $s$ is
$s^*=1/2$, because then $\Delta_{1-s}(P_{XX^\prime})=\Delta_s(P_{XX^\prime})$.
Thus, the r.h.s.\ of eq.\ (\ref{csiszarstyle}) is equivalent to
$$\inf_{\hat{R}< R}\inf_{\{P_{XX^\prime}:~
D(P_{XX^\prime}\|Q\times Q)\le 2\hat{R}\}}
\frac{\Delta_{1/2}(P_{XX^\prime})+\hat{R}}{R-\hat{R}}.$$
Now,
\begin{eqnarray}
& &\inf\{\Delta_{1/2}(P_{XX^\prime}): D(\tilde{P}_{XX^\prime}\|Q\times Q)
\le 2\hat{R}\}\nonumber\\
&=&\inf_{P_{XX^\prime}}\sup_{\rho\ge 0}
\left[\sum_{x,x^\prime}P_{XX^\prime}(x,x^\prime)
d_{1/2}(x,x^\prime)+
\rho\left(\sum_{x,x^\prime}P_{XX^\prime}(x,x^\prime)
\log\frac{P_{XX^\prime}(x,x^\prime)}{Q(x)Q(x^\prime)}-
2\hat{R}\right)\right]\nonumber\\
&=&\sup_{\rho\ge 0}\inf_{P_{XX^\prime}}\left[
\rho\cdot\sum_{x,x^\prime}P_{XX^\prime}(x,x^\prime)\log
\frac{P_{XX^\prime}(x,x^\prime)}
{Q(x)Q(x^\prime)2^{-d_{1/2}(x,x^\prime)/\rho}}-2\rho\hat{R}\right]\nonumber\\
&=&\sup_{\rho\ge
0}\left\{-\rho\log\left[\sum_{x,x^\prime}Q(x)Q(x^\prime)
2^{-d_{1/2}(x,x^\prime)/\rho}\right]-2\rho\hat{R}\right\}\nonumber\\
&=&\sup_{\rho\ge
0}\left\{-\rho\log\left[\sum_{x,x^\prime}Q(x)Q(x^\prime)
\left(\sum_y\sqrt{W(y|x)W(y|x^\prime)}\right)^{1/\rho}\right]-
2\rho\hat{R}\right\}\nonumber\\
&=&\sup_{\rho\ge 0}[E_{\mbox{\tiny x}}(\rho,Q)-2\rho\hat{R}],
\end{eqnarray}
and so,
\begin{eqnarray}
\label{almostdone}
\calE_{\mbox{\tiny trtc}}(R,Q)&\ge&\inf_{\hat{R} < R}\sup_{\rho\ge 0}\frac{E_{\mbox{\tiny
x}}(\rho,Q)-(2\rho-1)\hat{R}}{R-\hat{R}}\nonumber\\
&\ge&\inf_{\hat{R} < R}
\frac{E_{\mbox{\tiny x}}(\rho_{\mbox{\tiny trtc}}(R),Q)-(2\rho_{\mbox{\tiny
trtc}}(R)-1)\hat{R}}{R-\hat{R}}\nonumber\\
&=&\inf_{\hat{R} < R}\frac{(2\rho_{\mbox{\tiny
trtc}}(R)-1)R-(2\rho_{\mbox{\tiny trtc}}(R)-1)
\hat{R}}{R-\hat{R}}\nonumber\\
&=&2\rho_{\mbox{\tiny trtc}}(R)-1\nonumber\\
&=&\frac{E_{\mbox{\tiny x}}(\rho_{\mbox{\tiny trtc}}(R),Q)}{R}\nonumber\\
&=&E_{\mbox{\tiny trtc}}(R,Q).
\end{eqnarray}
Formally, this proves Theorem 1, but as a final remark, to complete the
picture, we also argue that the there is no loss of tightness in the passage
from the right--hand side of the first line of eq.\ (\ref{almostdone})
to $E_{\mbox{\tiny trtc}}(R,Q)$. This follows from the
following matching upper bound on the first line of (\ref{almostdone}).
Let $\tilde{R}$ be such that the maximizer of
$E_{\mbox{\tiny
x}}(\rho,Q)-(2\rho-1)\tilde{R}$ is $\rho_{\mbox{\tiny trtc}}(R)$. This is feasible due
to the concavity of $E_{\mbox{\tiny x}}(\rho,Q)$ in $\rho$ \cite[Theorem
3.3.2]{VO79},
$$\tilde{R}=\frac{1}{2}\cdot\frac{\partial E_{\mbox{\tiny
x}}(\rho,Q)}{\partial\rho}\bigg|_{\rho=\rho_{\mbox{\tiny trtc}}(R)}\le
\frac{E_{\mbox{\tiny x}}(\rho_{\mbox{\tiny trtc}}(R),Q)}{2\rho_{\mbox{\tiny
trtc}}(R)}\le \frac{E_{\mbox{\tiny
x}}(\rho_{\mbox{\tiny trtc}}(R),Q)}{2\rho_{\mbox{\tiny trtc}}(R)-1}=R.$$ 
Thus,
\begin{eqnarray}
\inf_{\hat{R} < R}\sup_{\rho\ge 0}
\frac{E_{\mbox{\tiny x}}(\rho,Q)-(2\rho-1)\hat{R}}{R-\hat{R}}
&\le&\sup_{\rho\ge 0}
\frac{E_{\mbox{\tiny x}}(\rho,Q)-(2\rho-1)\tilde{R}}{R-\tilde{R}}\nonumber\\
&=&\frac{E_{\mbox{\tiny x}}(\rho_{\mbox{\tiny trtc}}(R),Q)-(2\rho_{\mbox{\tiny
trtc}}(R)-1)\tilde{R}}{R-\tilde{R}}\nonumber\\
&=&\frac{(2\rho_{\mbox{\tiny trtc}}(R)-1)R-(2\rho_{\mbox{\tiny trtc}}(R)-1)
\tilde{R}}{R-\tilde{R}}\nonumber\\
&=&2\rho_{\mbox{\tiny trtc}}(R)-1\nonumber\\
&=&\frac{E_{\mbox{\tiny x}}(\rho_{\mbox{\tiny trtc}}(R),Q)}{R}\nonumber\\
&=&E_{\mbox{\tiny trtc}}(R,Q).
\end{eqnarray}

\section{Discussion}
\label{dis}

Several comments are in order concerning Theorem \ref{thm} and its proof.\\

\begin{figure}[ht]
\vspace*{1cm}
\hspace*{2cm}\input{graphical.pstex_t}
\caption{\small Graphical representation of $E_{\mbox{\tiny trtc}}(R,Q)$ and
$E_{\mbox{\tiny cex}}(R,Q)$.}
\label{graphical}
\end{figure}
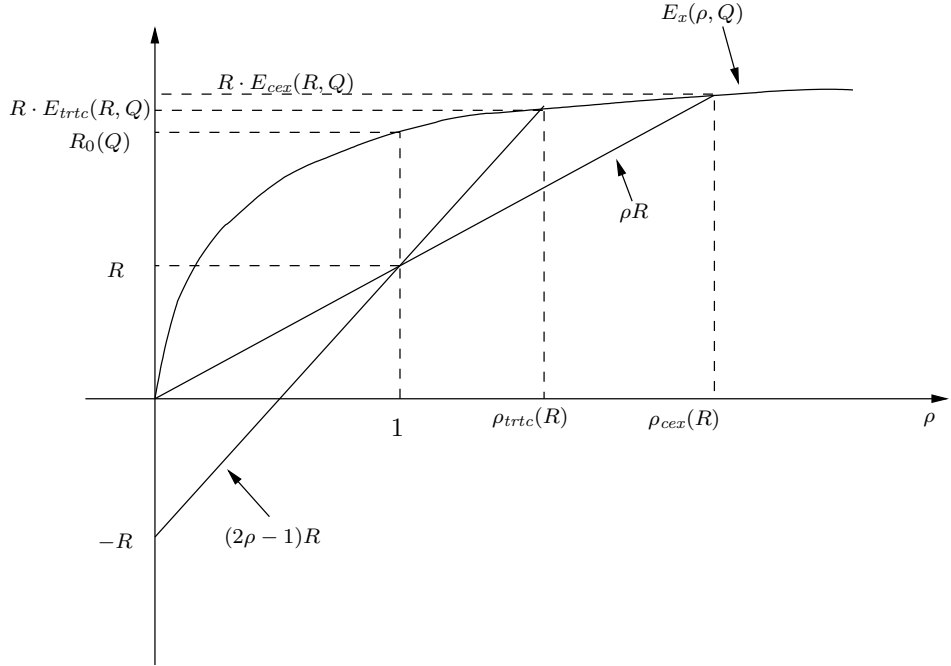

\noindent
{\bf Relations among the exponents.}
It is easy to see 
that $E_{\mbox{\tiny trtc}}(0)$ is equal to the zero--rate
expurgated exponent, 
$E_{\mbox{\tiny
ex}}(0,Q)=E_{\mbox{\tiny cex}}(0,Q)=\lim_{\rho\to\infty}E_{\mbox{\tiny
x}}(\rho,Q)$,
and that for all $R < R_0(Q)$,
$$E_{\mbox{\tiny rtc}}(R,Q)=\frac{R_0(Q)}{R}\le E_{\mbox{\tiny trtc}}(R,Q)
\le E_{\mbox{\tiny cex}}(R,Q).$$
In other words, the typical random trellis code exponent
is between the convolutional coding random coding exponent
and the convolutional coding expurgated exponent. This is parallel to the
ordering among the corresponding the block code exponents  \cite{trc}. These relations are
displayed graphically in
Fig.\ \ref{graphical}, where the concave curve of $E_{\mbox{\tiny
x}}(\rho,Q)$ is plotted as a function of $\rho$, along with the straight
lines, $\rho R$ and $(2\rho-1)R$. For $\rho=1$, we have $E_{\mbox{\tiny
x}}(1,Q)=E_0(1,Q)=R_0(Q)$. The
straight lines $\rho R$ and $(2\rho-1)R$ intersect at the point $(1,R)$, which is below the point
$(1,R_0(Q))$ on the curve (as $R$ is assumed smaller than $R_0(Q)$). 
The straight lines $\rho R$ and $(2\rho-1)R$ meet the curve
$E_{\mbox{\tiny x}}(\rho,Q)$ at the points
$(\rho_{\mbox{\tiny cex}}(R),R\cdot E_{\mbox{\tiny cex}}(R,Q))$ and
$(\rho_{\mbox{\tiny trtc}}(R),R\cdot E_{\mbox{\tiny trtc}}(R,Q))$,
respectively. As can be seen, $R\cdot E_{\mbox{\tiny cex}}(R,Q))\ge R\cdot E_{\mbox{\tiny
trtc}}(R,Q))\ge R_0(Q)$.\\

\noindent
{\bf Properties of the typical random trellis codes.} For
typical randomly selected trellis codes, we are able to characterize the features 
that make them achieve $E_{\mbox{\tiny trtc}}(R,Q)$. This is, in fact, spelled out
explicitly in the definition of the subset of typical codes, $\calT_k$. We know that
for these codes,
joint types that correspond to empirical distributions that are too far from
$Q\times Q$ (e.g., those that exhibit too strong empirical dependency between
the incorrect path and the correct one), are not populated.
For the other types, we know the distance spectrum, or more
precisely, the population profile of the various joint types.\\

\noindent
{\bf Dominant error events.} In the process of proving Theorem \ref{thm} in
Section \ref{proof},
we have seen also alternative forms of the error
exponent expression, like
the Csisz\'ar--style expression (\ref{csiszarstyle}). While this 
expression may not be easier to
calculate numerically (due to the nested optimizations involved), 
it is nevertheless useful for gaining some insight. We
learn the following from the first part of the derivation:
the error probability is dominated by a sub--exponential number of incorrect
paths whose joint empirical distribution with the correct path is given by
\begin{equation}
\label{pxxp}
P_{XX^\prime}^*(x,x^\prime)=\frac{Q(x)Q(x^\prime)2^{-d_{1/2}(x,x^\prime)/\rho}}
{\sum_{\hat{x},\tilde{x}}Q(\hat{x})Q(\tilde{x})
2^{-d_{1/2}(\hat{x},\tilde{x})/\rho}}
\end{equation}
and whose total unmerged length, $k+\ell$ (a.k.a.\ the
critical length), spans 
$$k+k\theta(D(P_{XX'}^*\|Q\times Q))=kR/[2R-D(P_{XX'}^*\|Q\times Q)]$$
branches.\footnote{Interestingly, this is different from the total critical length that dominates
ordinary average error probability, which for $R < R_0$, is $k$ branches long
\cite[Theorem 5.5.1]{VO79}.} The error exponent expression
(\ref{csiszarstyle}) is therefore essentially the same as that of a
zero--rate\footnote{The zero rate is because of the sub--exponential number of
dominant incorrect paths.} block 
code of block length $K/[2R-D(P_{XX'}^*\|Q\times Q)]$, where the competing
trellis paths are at normalized Bhattacharyya distance $\Delta_{1/2}(P_{XX'}^*)$
from the correct path, hence
the product, $\Delta_{1/2}(P_{XX'}^*)/[2R-D(P_{XX'}^*\|Q\times Q)]$.
For time--varying convolutional codes over the binary--input,
output--symmetric channel,
better performance is obtained (as discussed above) as one obtains \cite[Corollary 5.3.1]{VO79}, 
$$E_{\mbox{\tiny cex}}(R,Q)=\frac{\log Z}{\log(2^{1-R}-1)},$$
with $Z=\sum_y\sqrt{W(y|0)W(y|1)}$, which has the simple interpretation of the
Costello lower bound on the free distance \cite{Costello74} multiplied by the
corresponding Bhattacharyya bound (see also \cite[p.\ 1652]{ZSSHJ99}).
In other words, the typical time--varying convolutional code achieves the Costello bound.
Note that the parameter $\rho$ in (\ref{pxxp}) controls the
similarity (and hence the dominant distance) between $P_{XX^\prime}^*$ and the
product distribution $Q\times Q$. When $\rho$ is very large (at low
rates), the dominant distance is large and when $\rho$ is very small (low
rates), the distance is very small.\\

\begin{figure}[h!t!b!]
\centering
\includegraphics[width=8.5cm, height=8.5cm]{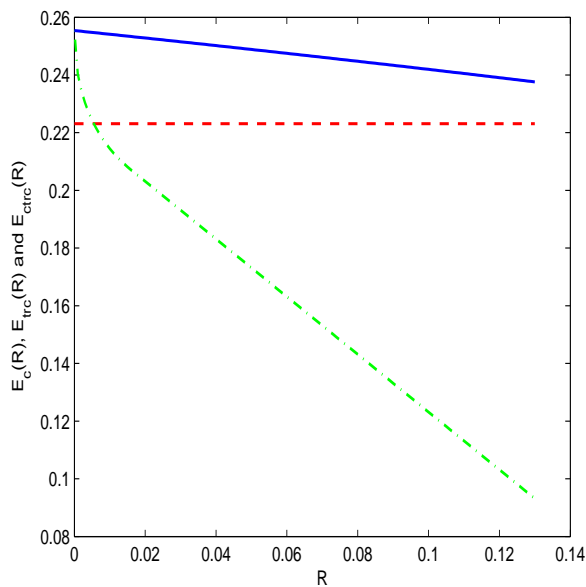}
\caption{The functions $E_{\mbox{\tiny trc}}(R,Q)$ \cite{trc} of general (unstructured) random block
codes (green dashed curve
$-\cdot-$), 
$R\cdot E_{\mbox{\tiny rtc}}(R)\equiv R_0$ of the 
random convolutional coding exponent (red dashed curve $---$) and
$R\cdot E_{\mbox{\tiny trcc}}(R)=E_{\mbox{\tiny
x}}(\rho_{\mbox{\tiny cex}}(R),Q)$ of the typical random coding convolutional code 
(blue solid curve), all in the range $[0,R_{\mbox{\tiny crit}}]$ for the
binary symmetric channel with
crossover parameter
$p=0.1$, where $R_0=0.2231$ and $R_{\mbox{\tiny
crit}}=0.1308$.
All rates are in units of nats/channel--use.}
\label{graph3}
\end{figure}

\noindent
{\bf A numerical example.} In \cite[Chap. 5]{VO79}, there is a comparison of the
performance--complexity trade-off between unstructured block codes and convolutional codes,
where the performance is measured according to the traditional random coding error
exponents. As explained therein, the idea is that for block codes of length $N$ and rate $R$, the
complexity is $G=2^{NR}$ and the error probability is
exponentially $2^{-NE_{\mbox{\tiny block}}(R)}=G^{-E_{\mbox{\tiny
block}}(R)/R}$. For convolutional codes, decoded by the Viterbi algorithm, the
complexity is about $G=2^K$ and the error probability decays like
$2^{-KE_{\mbox{\tiny conv}}(R)}=G^{-E_{\mbox{\tiny conv}}(R)}$, and so, it
makes sense to compare $E_{\mbox{\tiny
block}}(R)/R$ with $E_{\mbox{\tiny conv}}(R)$, or more conveniently, to compare
$E_{\mbox{\tiny block}}(R)$ with $R\cdot E_{\mbox{\tiny conv}}(R)$.
It is interesting to conduct a similar comparison when the
performance of both classes of codes is measured according to error exponents
of the typical random codes. In Fig.\ \ref{graph3}, this is done for the
binary symmetric channel with crossover parameter $p=0.1$ and the uniform random coding
distribution. For reference, the ordinary random coding exponent of
convolutional codes, $R\cdot E_{\mbox{\tiny rtc}}(R,Q)\equiv R_0(Q)$, is also
plotted in the displayed range of rates. As can be seen, the
typical code exponent of the ensemble of time--varying convolutional codes
is much larger than that of block codes for the same
decoding complexity.

\section{Channels with Memory and Mismatch}

In this section, we extend our main results in two directions at the same time.
The first direction is that instead of 
assuming memoryless channels, we now allow channels that memorize a finite
number of the most recent past inputs, with the clear motivation of channels with
intersymbol interference (see also \cite[Sect.\ 5.8]{VO79}). 
For the sake of simplicity, we consider the case where the memory contains the
one most recent past input only, in other words, the channel model (\ref{channel}) is
replaced by
\begin{equation}
\label{isichannel}
\mbox{Pr}\{Y_1=y_1,Y_2=y_2,\ldots,Y_r=y_r|X_0=x_0,X_1=x_1,\ldots,X_r=x_r\}=
\prod_{t=1}^rW(y_t|x_t,x_{t-1}).
\end{equation}
The extension to any fixed number $p$ of the 
most recent past inputs is conceptually straightforward
by redefining the channel input at time $t$ as
$\bar{x}_t=(x_t,\ldots,x_{t-p+1})$ and taking into account that in the
sequence $\{\bar{x}_t\}$ not all $(J^p)^2$
state transitions $\bar{x}_t\to\bar{x}_{t+1}$ are allowed, but only those in
which the two states are
consistent with each other. Using this transformation, we are back to the
model (\ref{isichannel}), except that $\{x_t\}$ are replaced by $\{\bar{x}_t\}$.
The other direction of extension is that we allow mismatch. The decoding
metric is assumed to be $\prod_t \tW(y_t|x_t,x_{t-1})$ for some channel
$\tW$ that may differ from $W$. To avoid further complications, 
the ensemble of time--varying trellis codes
continues to be
defined exactly as in Section \ref{npbo} (without any attempt at introducing
memory). These model assumptions are motivated by the
facts that: (i) they are practically relevant, and (ii) the Viterbi algorithm is
still implementable, although the number of states is now larger than before.
In the remaining part of this section, we will not repeat all the
derivations of Section \ref{proof}, but only highlight the differences and the
state the results.

The first basic difference, relative to the derivation in Section \ref{proof},
is associated the pairwise error probability: given the correct trellis path $\bx$
and a competing path $\bx'$, both of length $n(k+\ell)$ channel uses, 
the pairwise average error probability is upper bounded using the
Chernoff bound as follows:
\begin{eqnarray}
\bar{P}_{\mbox{\tiny e}}(\bx\to\bx')&\le&\sum_{\bx,\bx'}Q(\bx)Q(\bx')
\cdot\min_{s\ge 0}\sum_{\by}W(\by|\bx)\cdot
\left[\frac{\tW(\by|\bx')}{\tW(\by|\bx)}\right]^s\nonumber\\
&=&\sum_{\bx,\bx'}Q(\bx)Q(\bx')
\cdot\min_{s\ge
0}\sum_{\by}\prod_{t=1}^{n(k+\ell)}W(y_t|x_t,x_{t-1})
\tW^{1-s}(y_t|x_t,x_{t-1})\tW^s(y_t|x_t',x_{t-1}')\nonumber\\
&=&\sum_{\bx,\bx'}Q(\bx)Q(\bx')
\cdot\min_{s\ge
0}\prod_{t=1}^{n(k+\ell)}\sum_{y_t}W(y_t|x_t,x_{t-1})\tW^{1-s}(y_t|x_t,x_{t-1})
\tW^s(y_t|x_t',x_{t-1}')\nonumber\\
&=&\sum_{\bx,\bx'}Q(\bx)Q(\bx')
\cdot\min_{s\ge 0}\exp_2\left\{-\sum_{t=1}^{n(k+\ell)}d_s(x_t,x_{t-1};x_t',x_{t-1}')\right\}\nonumber\\
&=&\sum_{\bx,\bx'}Q(\bx)Q(\bx')
\cdot\exp_2\left\{-\max_{s\ge 0}\sum_{t=1}^{n(k+\ell)}d_s(x_t,x_{t-1};x_t',x_{t-1}')\right\}
\end{eqnarray}
where we have defined
\begin{equation}
d_s(x,x_-;x',x_-')=-\log\left[\sum_y W(y|x,x_-)\tW^{1-s}(y|x,x_-)\tW^s(y|x',x_-')\right],
~~~~x,x_-,x',x_-'\in\calX.
\end{equation}
Note that here, it is no longer necessarily true that the optimal choice of
$s$ is $s=1/2$, as the symmetry properties that were valid in the memoryless
matched case of Section \ref{proof}, do not continue to hold here, in general. To make the
derivation more tractable, in the sequel,
we interchange the optimization over $s$ with the summation over
$\{\bx,\bx^\prime\}$, at the possible risk of losing exponential
tightness.\footnote{
Of course, one may always select $s=1/2$, as in Section \ref{proof}, and
then Theorem \ref{thm} will still be obtained as a special case.}
The expression 
$\sum_{t=1}^{n(k+\ell)}d_s(x_t,x_{t-1};x_t',x_{t-1}')$ depends on $(\bx,\bx')$
only via their joint ``Markov type'', defined by the joint empirical distribution,
\begin{equation}
\hat{P}_{XX'X_-X_-'}(x,x',x_-,x_-')=\frac{1}{k+\ell}
\sum_{t=1}^{n(k+\ell)}\calI\{x_t=x,x_t'=x',x_{t-1}=x_-,x_{t-1}'=x_-'\},
\end{equation}
ignoring edge effects. Let us denote
\begin{equation}
\Delta_s(\hat{P}_{XX'X_-X_-'})=\sum_{x,x',x_-,x_-'}\hat{P}_{XX'X_-X_-'}(x,x',x_-,x_-')
d_s(x,x_-;x',x_-').
\end{equation}
Using the extension of the method of 
types to Markov types (see, e.g., 
\cite[Sect.\ VII.A]{Csiszar98}, \cite{DLS81}, \cite[Sect.\ 3.1]{DZ93},
\cite{Natarajan85}), we find that
\begin{eqnarray}
\bar{P}_{\mbox{\tiny e}}(\bx\to\bx')&\lexe&\min_{s\ge 0}
\max_{\hat{P}_{XX'X_-X_-'}}
\exp\bigg\{n(k+\ell)\bigg[\hat{H}(X,X'|X_-,X_-')-\hat{H}(X,X')-\nonumber\\
& &D(\hat{P}_{XX'}\|Q\times Q)-\Delta_s(\hat{P}_{XX'X_-X_-'})\bigg]\bigg\}\nonumber\\
&=&\exp\bigg\{-n(k+\ell)\max_{s\ge 0}\min_{\hat{P}_{XX'X_-X_-'}}\bigg[
D(\hat{P}_{XX'|X_-X_-'}\|Q\times Q|\hat{P}_{X_-X_-'})+\nonumber\\
& &\Delta_s(\hat{P}_{XX'X_-X_-'})\bigg]\bigg\},
\end{eqnarray}
where $\hat{H}(X,X'|X_-,X_-')$ is the empirical conditional entropy of
$(X,X')$ given $(X_-,X_-')$, derived from $\hat{P}_{XX'X_-X_-'}$, 
\begin{equation}
D(\hat{P}_{XX'|X_-X_-'}\|Q\times Q|\hat{P}_{X_-X_-'})\dfn\sum_{x,x_-,x',x_-'}
\hat{P}_{XX'X_-X_-'}(x,x',x_-,x_-')\log
\frac{\hat{P}_{XX'|X_-X_-'}(x,x'|x_-,x_-')}{Q(x)Q(x')},\nonumber\\
\end{equation}
$\hat{P}_{XX'|X_-X_-'}$ being the conditional distribution induced by
$\hat{P}_{XX'X_-X_-'}$, and
the minimization over $\{\hat{P}_{XX'X_-X_-'}\}$ is confined to joint
distributions where the marginals of $(X,X')$ and $(X_-,X_-')$ are the same.
Repeating the same steps as in Section \ref{proof}, and assuming that
\begin{equation}
\label{RoQ}
R < R_0(Q)=\max_{s\ge 0}\min_{P_{XX'X_-X_-'}}[D(P_{XX'|X_-X_-'}\|Q\times
Q|P_{X_-X_-'})+\Delta_s(P_{XX'X_-X_-'})],
\end{equation}
the resulting error exponent of the typical random trellis 
code is lower bounded by
\begin{equation}
\max_{s\ge 0}\min_{\hat{R} < R}
\min_{\{\hat{P}_{XX'X_-X_-'}:~D(\hat{P}_{XX'|X_-X_-'}\|Q\times
Q|\hat{P}_{X_-X_-'})\le 2\hat{R}\}}
\frac{\Delta_s(\hat{P}_{XX'X_-X_-'})+\hat{R}}{
R-\hat{R}}.
\end{equation}
As for the inner--most minimization,
let us define the functions
\begin{equation}
F_s(d)=\min\{D(\hat{P}_{XX'|X_-X_-'}\|Q\times
Q|\hat{P}_{X_-X_-'}):~\Delta_s(\hat{P}_{XX'X_-X_-'})\le d\}
\end{equation}
and 
\begin{equation}
G_s(2\hat{R})=\min\{\Delta_s(\hat{P}_{XX'X_-X_-'}):~D(\hat{P}_{XX'|X_-X_-'}\|Q\times
Q|\hat{P}_{X_-X_-'})\le 2\hat{R}\}.
\end{equation}
From large deviations theory \cite[Sect.\ 3.1]{DZ93}, 
we know that an alternative expression for $F_s(d)$ is given by
\begin{equation}
F_s(d)=\sup_{r\ge 0}[G_s(r)-rd],
\end{equation}
where $G_s(r)=-\log\lambda_s(r)$, $\lambda_s(r)$ being the Perron--Frobenius
eigenvalue of the $J^2\times J^2$ matrix 
$$A_s(r)=\{Q(x)Q(x^\prime)e^{-rd_s(x,x_-;x',x_-')}\}$$
whose rows and columns are indexed by the pairs $(x,x')$ and $(x_-,x_-')$,
respectively.\footnote{
This equivalence between the two forms of $F_s(d)$ follows from the fact that
they are both expressions of the large deviations rate function \cite[Sect.\
3.1]{DZ93} of the
probability of the event
$\{\sum_{t=1}^Nd_s(X_t,X_{t-1};X_t^\prime,X_{t-1}^\prime)\le Nd\}$, where
$\{X_t\}$ and $\{X_t^\prime\}$ are independent i.i.d.\ processes, both
governed by $Q$.}
Thus, given $d$,
$$\Delta_s(\hat{P}_{XX'X_-X_-'})\le d~\mbox{implies}~D(\hat{P}_{XX'|X_-X_-'}\|Q\times
Q|\hat{P}_{X_-X_-'})\ge F_s(d).$$
Equivalently, given that $2\hat{R}=F_s(d)$, 
$$D(\hat{P}_{XX'|X_-X_-'}\|Q\times Q|\hat{P}_{X_-X_-'})\le
2\hat{R}~\mbox{implies}~
\Delta_s(\hat{P}_{XX'X_-X_-'})\ge F_s^{-1}(2\hat{R}).$$
But
\begin{equation}
F_s^{-1}(2\hat{R})=\sup_{r\ge 0}\frac{G_s(r)-\hat{R}}{r}=\sup_{\rho\ge 0}
[\rho G_s(1/\rho)-2\rho\hat{R}],
\end{equation}
and so, similarly as in Section \ref{proof}, 
the error exponent of the typical random trellis code is lower bounded by
\begin{equation}
\sup_{s\ge 0}\inf_{\hat{R} < R}
\sup_{\rho\ge 0}\frac{\rho G_s(1/\rho)-(2\rho-1)\hat{R}}{R-\hat{R}}
=\sup_{s\ge 0}\frac{\rho_{R,s}G_s(1/\rho_{R,s})}{R},
\end{equation}
where $\rho_{R,s}$ is the solution to the equation $(2\rho-1)R=\rho G_s(1/\rho)$.
Note that $\rho G_s(1/\rho)$ is an extension of $E_{\mbox{\tiny x}}(\rho,Q)$ to a channel
with both memory and mismatch. Using similar considerations, it is easy to see
that $R_0(Q)$ of eq.\ (\ref{RoQ}) is equal to $\sup_{s\ge 0}G_s(1)$.

Referring to the comment on the extension to channels
with memory of the $p$ most recent past channel
inputs (see the introductory paragraph of this section), 
the only difference is that in such a case, the matrix $A_s(r)$ has larger dimensions,
$J^{2p}\times J^{2p}$, but it is rather sparse: all entries vanish
except those where both pairs $(x,x_-)$ and
$(x^\prime,x_-^\prime)$ are consistent.

\end{document}

%% file: trelliscode.pstex_t
\begin{picture}(0,0)%
\includegraphics{trelliscode.pstex}%
\end{picture}%
\setlength{\unitlength}{4144sp}%
\begingroup\makeatletter\ifx\SetFigFont\undefined%
\gdef\SetFigFont#1#2#3#4#5{%
  \reset@font\fontsize{#1}{#2pt}%
  \fontfamily{#3}\fontseries{#4}\fontshape{#5}%
  \selectfont}%
\fi\endgroup%
\begin{picture}(6086,2437)(166,-1873)
\put(1702,-341){\makebox(0,0)[lb]{\smash{{\SetFigFont{11}{13.2}{\rmdefault}{\mddefault}{\itdefault}{$f_t$}%
}}}}
\put(1967,-799){\makebox(0,0)[lb]{\smash{{\SetFigFont{11}{13.2}{\rmdefault}{\mddefault}{\itdefault}{$\bx_t$}%
}}}}
\put(3005,262){\makebox(0,0)[lb]{\smash{{\SetFigFont{11}{13.2}{\rmdefault}{\mddefault}{\itdefault}{$\bu_{t-k+1}$}%
}}}}
\put(181,238){\makebox(0,0)[lb]{\smash{{\SetFigFont{11}{13.2}{\rmdefault}{\mddefault}{\itdefault}{$\bu_t$}%
}}}}
\put(4236,-1354){\makebox(0,0)[lb]{\smash{{\SetFigFont{11}{13.2}{\rmdefault}{\mddefault}{\itdefault}{$\by_t$}%
}}}}
\put(5853,-1354){\makebox(0,0)[lb]{\smash{{\SetFigFont{11}{13.2}{\rmdefault}{\mddefault}{\itdefault}{$\hat{\bu}_t$}%
}}}}
\put(3150,-1499){\makebox(0,0)[lb]{\smash{{\SetFigFont{11}{13.2}{\rmdefault}{\mddefault}{\itdefault}{$W(\by_t|\bx_t)$}%
}}}}
\put(4960,-1523){\makebox(0,0)[lb]{\smash{{\SetFigFont{11}{13.2}{\familydefault}{\mddefault}{\updefault}{decoder}%
}}}}
\end{picture}%

%% file: graphical.pstex_t
\begin{picture}(0,0)%
\includegraphics{graphical.pstex}%
\end{picture}%
\setlength{\unitlength}{4144sp}%
\begingroup\makeatletter\ifx\SetFigFont\undefined%
\gdef\SetFigFont#1#2#3#4#5{%
  \reset@font\fontsize{#1}{#2pt}%
  \fontfamily{#3}\fontseries{#4}\fontshape{#5}%
  \selectfont}%
\fi\endgroup%
\begin{picture}(5652,4040)(616,-3672)
\put(1218,-1333){\makebox(0,0)[lb]{\smash{{\SetFigFont{8}{9.6}{\rmdefault}{\mddefault}{\itdefault}{$R$}%
}}}}
\put(2917,-2289){\makebox(0,0)[lb]{\smash{{\SetFigFont{10}{12.0}{\rmdefault}{\mddefault}{\itdefault}{$1$}%
}}}}
\put(6104,-2206){\makebox(0,0)[lb]{\smash{{\SetFigFont{8}{9.6}{\rmdefault}{\mddefault}{\itdefault}{$\rho$}%
}}}}
\put(1163,-2959){\makebox(0,0)[lb]{\smash{{\SetFigFont{8}{9.6}{\rmdefault}{\mddefault}{\itdefault}{$-R$}%
}}}}
\put(1872,-211){\makebox(0,0)[lb]{\smash{{\SetFigFont{8}{9.6}{\rmdefault}{\mddefault}{\itdefault}{$R\cdot E_{\mbox{\tiny cex}}(R,Q)$}%
}}}}
\put(986,-566){\makebox(0,0)[lb]{\smash{{\SetFigFont{8}{9.6}{\rmdefault}{\mddefault}{\itdefault}{$R_0(Q)$}%
}}}}
\put(3522,-2211){\makebox(0,0)[lb]{\smash{{\SetFigFont{8}{9.6}{\rmdefault}{\mddefault}{\itdefault}{$\rho_{\mbox{\tiny trtc}}(R)$}%
}}}}
\put(4531,209){\makebox(0,0)[lb]{\smash{{\SetFigFont{8}{9.6}{\rmdefault}{\mddefault}{\itdefault}{$E_{\mbox{\tiny x}}(\rho,Q)$}%
}}}}
\put(4284,-979){\makebox(0,0)[lb]{\smash{{\SetFigFont{8}{9.6}{\rmdefault}{\mddefault}{\itdefault}{$\rho R$}%
}}}}
\put(1919,-2926){\makebox(0,0)[lb]{\smash{{\SetFigFont{8}{9.6}{\rmdefault}{\mddefault}{\itdefault}{$(2\rho-1)R$}%
}}}}
\put(4456,-2221){\makebox(0,0)[lb]{\smash{{\SetFigFont{8}{9.6}{\rmdefault}{\mddefault}{\itdefault}{$\rho_{\mbox{\tiny cex}}(R)$}%
}}}}
\put(631,-365){\makebox(0,0)[lb]{\smash{{\SetFigFont{8}{9.6}{\rmdefault}{\mddefault}{\itdefault}{$R\cdot E_{\mbox{\tiny trtc}}(R,Q)$}%
}}}}
\end{picture}%